\documentclass[aps,prl,twocolumn,groupedaddress]{revtex4-2}
\usepackage{latexsym,amsmath,amsfonts,amssymb,amsthm,mathrsfs, mathtools}
\usepackage{graphicx}

\usepackage{caption, subcaption}
\usepackage{multirow}
\usepackage{tensor}

\usepackage{tikz}

\usepackage{hyperref}

\def\m{\mu}

\def\bbC{\mathbb{C}}

\def\beq#1\eeq{\begin{align}#1\end{align}}

\usepackage{enumitem}

\def\bbC{\mathbb{C}}

\def\beq#1\eeq{\begin{align}#1\end{align}}

\theoremstyle{definition}

\begin{document}

% Use the \preprint command to place your local institutional report
% number in the upper righthand corner of the title page in preprint mode.
% Multiple \preprint commands are allowed.
% Use the 'preprintnumbers' class option to override journal defaults
% to display numbers if necessary
%\preprint{}

%Title of paper
\title{Chiral algebra, Wilson lines, and the mixed Hodge structure of the Coulomb branch}

% repeat the \author .. \affiliation  etc. as needed
% \email, \thanks, \homepage, \altaffiliation all apply to the current
% author. Explanatory text should go in the []'s, actual e-mail
% address or url should go in the {}'s for \email and \homepage.
% Please use the appropriate macro foreach each type of information

% \affiliation command applies to all authors since the last
% \affiliation command. The \affiliation command should follow the
% other information
% \affiliation can be followed by \email, \homepage, \thanks as well.
\author{Yutong Li}
\author{Yiwen Pan}
%\email[]{Your e-mail address}
%\homepage[]{Your web page}
%\thanks{} %\altaffiliation{}
\affiliation{Department of Physics,  Sun Yat-Sen University, Guangzhou, Guangdong, China}

\author{Wenbin Yan}
%\email[]{Your e-mail address}
%\homepage[]{Your web page}
%\thanks{}
%\altaffiliation{}
\affiliation{Yau Mathematics Science Center, Tsinghua University, Beijing, China}

%Collaboration name if desired (requires use of superscriptaddress
%option in \documentclass). \noaffiliation is required (may also be
%used with the \author command).
%\collaboration can be followed by \email, \homepage, \thanks as well.
%\collaboration{}
%\noaffiliation

\date{\today}

\begin{abstract}
% insert abstract here
We find an intriguing relation between the chiral algebra and the mixed Hodge structure of the Coulomb branch of four-dimensional $\mathcal{N} = 2$ superconformal field theories. We identify the space of irreducible characters of the $\mathcal{N} = 4$ $SU(N)$ chiral algebra $\mathbb{V}[\mathcal{T}_{SU(N)}]$ by analytically computing the Wilson line Schur index and imposing modular invariance. We further establish a map from the characters of $\mathbb{V}[\mathcal{T}_{SU(N)}]$ to those of the $\mathcal{T}_{p, N}$ chiral algebra. We extract the pure part of the mixed Hodge polynomial $PH_c$ of the Coulomb branch compactified on a circle and prove that $PH_c$ encodes the representation theory of $\mathbb{V}[\mathcal{T}_{SU(N)}]$. We expect this to provide a new entry in the 4D mirror-symmetry framework.
\end{abstract}

% insert suggested keywords - APS authors don't need to do this
%\keywords{}

%\maketitle must follow title, authors, abstract, and keywords
\maketitle

\section{Introduction}

The rich structure of four-dimensional (4D) $\mathcal{N} = 2$ superconformal field theories (SCFTs) is encoded in a wide variety of physical observables. Surprisingly, these observables often exhibit hidden relations. In this Letter, we investigate an exact correspondence between two seemingly unrelated objects: the representation theory of the nonunitary associated chiral algebra \cite{Beem:2013sza}, which is tied, to some extent, to the Higgs branch of vacua, and the mixed Hodge polynomial associated with the Coulomb branch compactified on a circle \cite{hausel2008mixed}.

It is widely believed that non-vacuum modules of the associated chiral algebra $\mathbb{V}[\mathcal{T}]$ encode the BPS line and surface operators in the 4D theory $\mathcal{T}$ \cite{Cordova:2016uwk,Cordova:2017mhb,Bianchi:2019sxz,Nishinaka:2018zwq,Zheng:2022zkm}. The underlying physical picture is as follows: operators in the chiral algebra may participate in bulk-defect operator product expansions with defect Schur operators, thereby generating further defect Schur operators and giving rise to representations.

However, at the level of exactly computable physical observables, the correspondence is far from straightforward. The simplest instance is the identification of the Schur index $\mathcal{I}$ with the $\mathbb{V}[\mathcal{T}]$ vacuum character $\operatorname{ch}_0$. In \cite{Cordova:2017mhb,Nishinaka:2018zwq}, vortex-defect Schur indices were shown to reproduce the irreducible characters of the chiral algebra exactly. On the other hand, in a few examples, line-defect Schur indices were shown to be linear combinations of chiral algebra characters with nonconstant coefficients \cite{Cordova:2016uwk}. Schematically, the line index reads
\begin{equation}
  \mathcal{I}_L(q,b) = \sum_k R_k(q,b) \operatorname{ch}_k(q,b)\ .
\end{equation}
Here $q = e^{2\pi i \tau}$ denotes the standard parameter used to define a character or index, while $b$ collectively denotes the flavor fugacities. Because of the functional coefficients $R_k(q, b)$, the resulting index $\mathcal{I}_L$ is not a character of $\mathbb{V}[\mathcal{T}]$, whether irreducible or reducible, in the sense that it does not solve the required modular differential equations \cite{Beem:2017ooy,Arakawa:2016hkg,Zheng:2022zkm}. In fact, apart from a class of Argyres-Douglas theories and a few sporadic cases, the representation theory of $\mathbb{V}[\mathcal{T}]$ is entirely unknown because of the highly nonrational and logarithmic nature of these algebras, let alone its quantitative relation to 4D BPS nonlocal operators. Nonetheless, the existing observations strongly suggest that irreducible characters provide a complete basis for defect Schur indices.

Another crucial observable is the Coulomb 
branch of vacua of an $\mathcal{N} = 2$ SCFT. For a class-$\mathcal{S}$ theory of type $G$ engineered by a Riemann surface $\Sigma_{g, n}$, the Coulomb branch $\mathcal{M}_C$ of the 4D theory compactified on a circle is given by the hyperk\"{a}hler Hitchin moduli space $\mathcal{M}_H$ of $\Sigma_{g, n}$. Alternatively, the Coulomb branch can be described topologically as the character variety $\mathcal{M}$ of the Riemann surface $\Sigma_{g, n}$, namely, the space of homomorphisms from the fundamental group $\pi_1(\Sigma_{g, n})$ to the chosen gauge group $G$, modulo conjugation. The two descriptions respectively emphasize the holomorphic-symplectic and algebraic-topological features of the Coulomb branch $\mathcal{M}_C$. The conical singularity, together with the holomorphic-topological features of $\mathcal{M}_C$, gives rise to a \emph{mixed Hodge structure}, often expressed in terms of mixed Hodge polynomials $H_c$. A priori, this cohomological structure seems entirely independent of the chiral algebra $\mathbb{V}[\mathcal{T}]$.

The work presented in this Letter is twofold. First, we determine all the characters of the $\mathcal{N} = 4$ $SU(N)$ chiral algebra $\mathbb{V}[\mathcal{T}_{SU(N)}]$ and their modular properties. Second, we prove that the representation theory is encoded in the seemingly unrelated mixed Hodge polynomial of the Coulomb branch.

Concretely, we focus on Wilson line operators corresponding to arbitrary irreducible representations of $SU(N)$ and compute their line-defect indices analytically. We extract the space of irreducible characters of the chiral algebra and show that its basis elements are in one-to-one correspondence with the integer partitions $\boldsymbol{\ell}$ of $N$\footnote{Yongchao Lü independently conjectured such a relation between characters and partitions of $N$ \cite{Lu}.}. The logarithmic nature of the algebra implies the existence of logarithmic characters, which are constructed by imposing modular invariance on the character space $\mathcal{V}$.

We will prove that the structure of $\mathcal{V}$ is encoded in the Coulomb branch by computing the pure part of its mixed Hodge polynomial. Concretely, we find
\begin{equation}
  PH_c(\mathcal{M}, q) = \sum_{\boldsymbol{\ell} \in \mathcal{P}(N)} q^{\dim \mathcal{V}_{\boldsymbol{\ell}}-1} \ ,
\end{equation}
where $\mathcal{V}_{\boldsymbol{\ell}}$ contains all the logarithmic partners of the irreducible representations labeled by the partition $\boldsymbol{\ell}$.

% The chiral algebras $\mathbb{V}[\mathcal{T}]$ are in general logarithmic, leading to logarithmic representations and characters. To achieve modular invariance, we compute the $\Gamma^0(2)$-orbit of the $\mathcal{V}_\text{irred}$, which spans a much larger space $\mathcal{V}$. By picking suitable basis, we also compute the modular matrices $STS, T^2$, where $T^2$ has elegant closed form.

Our main finding may be considered a new entry in the four-dimensional mirror-symmetry framework \cite{2017arXiv170906142F,Dedushenko:2018bpp,Shan:2023xtw,Xie:2023pre,Shan:2024yas,Pan:2024hcz,Pan:2024epf}, revealing a deep relation between the representation-theoretic structure of the Higgs branch and the geometric structure of the Coulomb branch. Computing the mixed Hodge polynomial is generally much easier than computing the fixed-point varieties of the Hitchin moduli space, making it a powerful tool for studying 4D mirror symmetry and its relation to 3D mirror symmetry and symplectic duality.

\section{The chiral algebra and modules}

The associated chiral algebra $\mathbb{V}[\mathcal{T}_{SU(N)}]$ of the 4D $\mathcal{N} = 4$ $SU(N)$ theory $\mathcal{T}_{SU(N)}$ was proposed in \cite{Beem:2013sza} to be an $\mathcal{N} = 4$ super $\mathcal{W}$-algebra with $N - 1$ chiral primary generators of conformal weights $d_i/2$, where $d_i$ are the degrees of the Casimir invariants of $SU(N)$. In particular, the algebra contains the small 2D $\mathcal{N} = 4$ superconformal algebra.

In \cite{Bonetti:2018fqz}, a conjectural free-field realization of $\mathbb{V}[\mathcal{T}_{SU(N)}]$, associated with the Weyl group $S_N$ and based on $N - 1$ $bc\beta\gamma$ systems, was proposed. The conjecture was proven in \cite{Arakawa:2023cki}, and, more recently, the universal $\mathcal{W}$-algebra was shown to be unique \cite{Gaberdiel:2025eaf,Bonetti:2025kan}.

The representation theory of the simplest case, $N = 2$, was studied in \cite{Adamovic:2014lra}. There are two irreducible representations: the vacuum representation and the representation $M$, constructed as the quotient of one $bc\beta \gamma$ system by the vacuum. This $bc\beta \gamma$ system coincides with the one in \cite{Bonetti:2018fqz} and is a reducible but indecomposable representation of $\mathbb{V}[\mathcal{T}_{SU(2)}]$. The existence of only two irreducible representations can also be established by determining the two-dimensional solution space of the system of flavored modular linear differential equations (MLDEs) \cite{Pan:2021ulr,Zheng:2022zkm} associated with the three null states in the chiral algebra \cite{Beem:2017ooy}.

For $N > 2$, the situation is less clear. In the following, we identify the space of characters by computing the Wilson-line defect index of $\mathcal{T}_{SU(N)}$ for all $N \ge 2$.

\section{The Wilson line index}

There are BPS line defects that preserve the four supercharges used to define the Schur index \cite{Cordova:2016uwk}. These nonlocal operators can be inserted as a half-line or a full line passing through the origin of four-dimensional Euclidean spacetime. The simplest examples are Wilson line operators coupled to gauge fields in a Lagrangian theory. The corresponding defect Schur index can be written as a contour integral with the insertion of the appropriate Lie-group character of the representation carried by the half or full Wilson line \cite{Gang:2012yr}.

We consider $\mathcal{N} = 4$ $SU(N)$ super Yang-Mills theory. The (unnormalized) Schur index in the presence of a half Wilson line $W(Y)$ transforming under the $SU(N)$ representation $\mathcal{R}_Y$ (corresponding to the Young diagram $Y$) is given by the multivariate contour integral
\begin{equation}
  \mathcal{I}_{W(Y)}(q, b)
  = y^{- \frac{N^2 - 1}{2}}\oint \bigg[{\prod_{A = 1}^{N - 1}\frac{da_A}{2\pi i a_A}}\bigg]
  \mathcal{Z}(a) s_Y(a) \ ,
\end{equation}
where $s_Y(a)$ denotes the Schur polynomial associated with the representation $\mathcal{R}_Y$, and the integrand
\begin{align}
   \mathcal{Z}(a)\coloneqq& \  \\
  \frac{(-1)^{N + 1}}{N!}& \ \vartheta_4(\mathfrak{b})\eta(\tau)^{3(N - 1)}
  \frac{\prod^N_{A, B = 1 | A \ne B} \vartheta_1(\mathfrak{a}_A - \mathfrak{a})}{\prod_{A, B = 1}^N\vartheta_4(\mathfrak{a}_A - \mathfrak{a}_B + \mathfrak{b})} \nonumber \ .
\end{align}
The contour integral of $\mathcal{Z}(a)$ reproduces the Schur index $\mathcal{I}_{SU(N)}(q,b)$ of the theory without a Wilson line. We use different fonts to distinguish multiplicative variables from their additive counterparts: $a_A = e^{2\pi i \mathfrak{a}_A}$, $b = e^{2\pi i \mathfrak{b}}$, $y = e^{2\pi i \mathfrak{y}}$, and $q = e^{2\pi i \tau}$. The integration variables $a_A$ are constrained by $a_1 \cdots a_N = 1$. The prefactor $y^{- \frac{N^2 - 1}{2}}$ captures the flavor central charge $k_\text{2d} = - \frac{N^2 - 1}{2}$ and is often omitted except when discussing modular transformations.

The ratio of the Jacobi $\vartheta$ functions is elliptic in all the variables $\mathfrak{a}_A$, \emph{i.e.}, invariant under shifts of $\mathfrak{a}_A$ by $1$ or $\tau$. The integral vanishes whenever the number of boxes $|Y|$ in $Y$ is not divisible by $N$. On the other hand, when $|Y|$ is divisible by $N$, the constraint becomes unnecessary: one may treat $a_1, \dots, a_N$ as $N$ independent integration variables and replace $s_Y(a) \to (a_1 \cdots a_N)^{-|Y|/N}s_Y(a)$ without changing the result.

Consequently, without loss of generality, we can focus on the unconstrained integrals of the form
\begin{equation}
  \mathcal{I}(\mathbf{n}) \coloneqq \oint \prod_{A = 1}^N\frac{da_A}{2\pi i a_A}\mathcal{Z}(a)a_1^{n_1} \cdots a_N^{n_N} \ ,
\end{equation}
where $\mathbf{n} = (n_1, \dots, n_N)$ satisfies $\sum_{A = 1}^N n_A = 0$. The Wilson line index $\mathcal{I}_{W(Y)}$ is simply a linear combination of $\mathcal{I}(\mathbf{n})$ with integer coefficients. Since the ordering of the $n_i$ is irrelevant, we may treat $\mathbf{n}$ as a multiset.

The integral can be computed exactly using an integration formula \cite{Guo:2023mkn,Pan:2021mrw}\footnote{Details, additional examples of the computations, and generalizations to other theories will be reported in a companion paper \cite{Li:2026}.}; alternatively, one may use the Fermi-gas method \cite{Hatsuda:2023iwi}. In particular, the original Schur index is given by $\mathcal{I} = \mathcal{I}(\mathbf{0})$. Each integration step generates a Jacobi-form factor from the residue, a rational-function factor in $b$ and $q^{1/2}$, and at most one Eisenstein-series factor. The final result takes the form
\begin{equation}
  \mathcal{I}(\mathbf{n}) = \sum_{\alpha} R_\alpha(\mathbf{n}) \mathcal{J}_\alpha \ ,
\end{equation}
where $R_\alpha(\mathbf{n})$ denotes a rational function of $b$ and $q^{1/2}$, and the quasi-Jacobi form $\mathcal{J}_\alpha$ encodes the remaining Jacobi form and products of Eisenstein series. We now give an explicit description of the complete basis $\{\mathcal{J}_\alpha\}$.

\section{\texorpdfstring{The quasi-Jacobi basis $\mathcal{J}_{\boldsymbol{\ell}}$}{}}

First we decompose the set $\mathbf{n} = \{n_1, \cdots, n_N\}$ into $w$ disjoint subsets $\mathbf{n} = \bigcup_{j = 1}^{w} \{n_{j1}, \cdots, n_{j \ell_j}\}$ such that each subset sums to zero while containing no proper zero-sum subset. Hence,
\begin{equation}
  \sum_{j = 1}^{w}\ell_j = N \quad \text{and} \quad \sum_{k = 1}^{\ell_j}n_{jk} = 0 \ .
\end{equation}
We call the data $\boldsymbol{\ell} = [\ell_1, \cdots, \ell_w]$ the \emph{cycle type} of $\mathbf{n}$ and $\ell_i$ \emph{parts}, $w = w(\boldsymbol{\ell})$ the \emph{width} of $\boldsymbol{\ell}$. Without loss of generality, we may assume $\ell_1 \geq \ell_2 \geq \cdots \geq \ell_w$. Another useful notation for $\boldsymbol{\ell}$ is $[1^{m_1}2^{m_2}\cdots N^{m_N}]$ with multiplicities $m_A$.

The integral $\mathcal{I}(\mathbf{n})$ can be written as a linear combination of a basis of \emph{quasi-Jacobi forms} $\mathcal{J}_{\boldsymbol{\ell}}$ of weight $w(\boldsymbol{\ell}) - 1$,
\begin{equation}\label{Wilson-line-sum}
  \mathcal{I}(\mathbf{n}) = \sum_{\boldsymbol{\ell}' \in \text{Children}{(\boldsymbol{\ell})}} R_\mathbf{n}(\boldsymbol{\ell}') \mathcal{J}_{\boldsymbol{\ell}'} \ .
\end{equation}
The coefficients $R_{\mathbf{n}}(\boldsymbol{\ell}')$ are rational functions of $b$ and $q^{1/2}$ whose explicit form is not important for this Letter. The set $\operatorname{Children}(\boldsymbol{\ell})$ is determined by $\boldsymbol{\ell}$: a child $\boldsymbol{\ell}'$ of $\boldsymbol{\ell}$ is obtained by merging parts $\ell_i$, with the only restriction that an even number of unit-valued parts $\ell_i = 1$ cannot be merged into a single part. We denote this relation by ``parent $\succeq$ child,'' with $\boldsymbol{\ell} \succeq \boldsymbol{\ell}$ holding trivially. The set $\operatorname{Children}(\boldsymbol{\ell})$ contains $\boldsymbol{\ell}$ and all its children. This sum structure arises because $\prod_A a_A^{n_A}$ changes at each integration step as some of the variables $a_A$ are evaluated at poles.

The quasi-Jacobi forms $\mathcal{J}_{\boldsymbol{\ell}}$ can be written in closed form,
\begin{align}
  & \ \mathcal{J}_{\boldsymbol{\ell}} = 
  y^{- \frac{N^2 - 1}{2}}\frac{\vartheta_4(\mathfrak{b})}{i^{N^2 - 1}\vartheta_\text{1 or 4}(N \mathfrak{b})} \widehat{\mathcal{J}}_{\boldsymbol{\ell}} \nonumber\\
  = & \ y^{- \frac{N^2 - 1}{2}}\frac{\vartheta_4(\mathfrak{b})}{i^{N^2 - 1}\vartheta_\text{1 or 4}(N \mathfrak{b})} \\
  & \ \sum_{\sigma \in \mathcal{P}_{w - 1}(\{1, ..., w\})} \prod_{i} \left({- ({k_i} - 1)! E_{{k_i}} \begin{bmatrix}
    (-1)^{|\ell_{\sigma_i}|}\\
    b^{|\ell_{\sigma_i}|}
  \end{bmatrix}}\right) \ , \nonumber
\end{align}
where $\vartheta_{1\text{~or~}4}$ denotes $\vartheta_1$ when $N$ is even and $\vartheta_4$ when $N$ is odd. The set $\mathcal{P}_{w - 1}(\{1, ..., w\})$ consists of partitions $\sigma$ of all $(w-1)$-element subsets of $\{1, ..., w\}$,
\begin{equation}
  \bigcup_i \{\sigma_{i1}\cdots\sigma_{ik_i}\}\subset \{1, 2, \cdots, w\} \ ,
\end{equation}
such that $k_1 + k_2 + \cdots = w - 1$. We define $|\ell_{\sigma_i}| \coloneqq \ell_{\sigma_{i1}} + \cdots + \ell_{\sigma_{ik_i}}$.

We expect the basis $\{\mathcal{J}_{\boldsymbol{\ell}}\}$ to span the full space of irreducible characters of $\mathbb{V}[\mathcal{T}]$. In particular, the vacuum character reads
\begin{equation}
  \operatorname{ch}_0(q, b) = \mathcal{I}_{SU(N)}(q, b) = \sum_{\text{odd} ~\boldsymbol{\ell}} R_\mathbf{0}(\boldsymbol{\ell}) \mathcal{J}_{\boldsymbol{\ell}} \ ,
\end{equation}
where an odd $\boldsymbol{\ell}$ is a partition with only odd parts, and the coefficients $R_\mathbf{0}(\boldsymbol{\ell})$ are rational numbers,
\begin{equation}
  R_\mathbf{0}(\boldsymbol{\ell})
  = \frac{1}{\prod_{A = 1}^N m_A!}
  \prod_{j = 1}^{w(\boldsymbol{\ell})}
    \bigg({\frac{(\ell_j - 2)!!}{(2 i)^{(\ell_j - 1)/2}}}\bigg)^2 \frac{1}{\ell_j!}\ ,
\end{equation}
using the notations $\boldsymbol{\ell} = [\ell_1, \cdots, \ell_w] \sim [1^{m_1} \cdots N^{m_N}]$.

\section{\texorpdfstring{Map to $\mathcal{T}_{p, N}$ characters}{}}

We can establish a map from $\{\mathcal{J}_{\boldsymbol{\ell}}\}$ to known characters in other theories. Starting from the strongly coupled Argyres-Douglas theories $D_p(SU(N))$, one can construct new 4D SCFTs $\mathcal{T}_{p, N}$ \cite{Kang:2021lic,Jiang:2024baj}. The 4D central charges of $\mathcal{T}_{p, N}$ satisfy $a_\text{4d} = c_\text{4d}$, as do those of the $\mathcal{T}_{SU(N)}$ theories. More importantly, the vacuum character of the $\mathcal{T}_{p,N}$ chiral algebra is a specialization of that of $\mathcal{T}_{SU(N)}$,
\begin{equation*}
  \mathcal{I}_{\mathcal{T}_{p,N}}(q)= q^{- \frac{c_\text{2d}[\mathcal{T}_{p,N}] - p c_\text{2d}[\mathcal{T}_{SU(N)}]}{24}}\mathcal{I}_{SU(N)}(q^p, q^{\frac{p}{2}-1}) \ .
\end{equation*}
For $p = 3$ and $N = 2$, this relation was reinterpreted as a vector-space isomorphism between the vacuum representations of the two chiral algebras \cite{Buican:2020moo}.

Interestingly, the same specialization
\begin{equation}\label{eq:TpN-limit}
  b \to q^{\frac{p}{2} - 1}, \quad q \to q^p \ ,
\end{equation}
extends to a map between the representation theories of $\mathbb{V}[\mathcal{T}_{SU(N)}]$ and $\mathbb{V}[\mathcal{T}_{p, N}]$, where all characters of the latter theory are solutions to an unflavored modular differential equation. We have verified for $N = 2, \dots, 7$ that, under the specialization (\ref{eq:TpN-limit}), all the quasi-Jacobi forms $\mathcal{J}_{\boldsymbol{\ell}}$ map to characters of the $\mathcal{T}_{p, N}$ chiral algebra. Concretely, under (\ref{eq:TpN-limit}), the $\mathcal{J}_{\boldsymbol{\ell}}$ give solutions to the unflavored MLDE. For example, when $N = 5$,
\begin{align}
  \mathcal{J}_{[3,1,1]} \xrightarrow{b \to q^{\frac{p}{2} - 1}, q \to q^p} & \ -3 E_2(\tau) \ ,\\
  \mathcal{J}_{[1^5]} \xrightarrow{b \to q^{\frac{p}{2} - 1}, q \to q^p} & \ 15 E_2(\tau)^2 - 30 E_4(\tau) \ , 
\end{align}
giving two nonlogarithmic solutions to the 25th-order MLDE for $\mathcal{T}_{2,5}$ \cite{Pan:2025gzh}. Alternatively, the specialization can be expressed in terms of modular transformations acting on the vacuum character of the $\mathcal{T}_{p, N}$ chiral algebra.

\section{Modular properties}

We consider the following modular transformations,
\begin{align}
  S: (\tau, \mathfrak{b}, \mathfrak{y}) \to (- \frac{1}{\tau}, \frac{\mathfrak{b}}{\tau}, \mathfrak{y} - \frac{\mathfrak{b}^2}{\tau}), \\
  T: (\tau, \mathfrak{b}, \mathfrak{y}) \to (\tau + 1, \mathfrak{b}, \mathfrak{y}) \ .
\end{align}
Under $\Gamma^0(2) \subset SL(2, \mathbb{Z})$, generated by $STS$ and $T^2$, the overall $\vartheta$-ratio factor in $\mathcal{J}_{\boldsymbol{\ell}}$ transforms as a weight-zero Jacobi form. The remaining piece $\widehat{\mathcal{J}}_{\boldsymbol{\ell}}$, a sum of products of Eisenstein series, is a weight-$(w-1)$ quasi-Jacobi form. Its $\Gamma^0(2)$ orbit spans a $w$-dimensional linear space $\mathcal{V}_{\boldsymbol{\ell}}$ of quasi-Jacobi forms containing logarithmic terms. Whenever $\boldsymbol{\ell} \ne \boldsymbol{\ell}'$, the corresponding spaces intersect only at the origin. One can choose a basis and explicitly compute the modular-transformation matrices. In particular, we choose $\widehat{\mathcal{J}}_{\boldsymbol{\ell}}$, $STS \widehat{\mathcal{J}}_{\boldsymbol{\ell}}$, and $(T^2 - 1)^k STS \widehat{\mathcal{J}}_{\boldsymbol{\ell}}$ for $k = 1, \cdots, w - 2$. The $T^2$ matrix can be written as
\begin{equation}
  T^2 = i^{N^2 - 1} X \begin{pmatrix}
1 & 0 & \dots & \dots & \dots & 0 \\
0 & 1 & 1 & 0 & \dots & 0 \\
\vdots & 0 & 1 & 1 & 0 & \vdots \\
\vdots & \vdots & 0 & \ddots & \ddots & 0 \\
\vdots & \vdots & \vdots & 0 & 1 & 1 \\
0 & 0 & 0 & \dots & 0 & 1
\end{pmatrix}X^{-1} \ ,
\end{equation}
where $X$ is a $w\times w$ matrix. The Jordan normal form of $T^2$ is the Jordan block $J_{i^{N^2 - 1}, w}$. Since each $\boldsymbol{\ell}$ can be viewed as a partition of the integer $N$, we obtain the full space
\begin{align}
  \mathcal{V} = \bigoplus_{\boldsymbol{\ell} \in \mathcal{P}(N)} \mathcal{V}{_{\boldsymbol{\ell}}}, \qquad
  \dim \mathcal{V} = \sum_{\boldsymbol{\ell} \in \mathcal{P}(N)} w(\boldsymbol{\ell}) \ .
\end{align}

\section{Example}

As an example, when $N = 2$, there are only two quasi-Jacobi forms
\begin{equation}
  \mathcal{J}_{[2]} = \frac{\vartheta_4(\mathfrak{b})}{\vartheta_1(2 \mathfrak{b})} \qquad
  \mathcal{J}_{[1, 1]} = \frac{\vartheta_4(\mathfrak{b})}{\vartheta_1(2 \mathfrak{b})}E_2\begin{bmatrix}
    -1\\
    b
  \end{bmatrix} \ .
\end{equation}
The index of any spin-$j$ Wilson line is spanned by this basis,
\begin{equation}
  \mathcal{I}_j = \delta_{j\in \mathbb{Z}}(\mathcal{J}_{[1,1]}
  - \frac{i}{2}\mathcal{J}_{[2]}\sum_{\substack{m = -j\\m\ne0}}^{+j} \frac{b^m - b^{-m}}{q^{m/2} - q^{- m /2}}
  ) \ .
\end{equation}
The two $\mathcal{J}$ precisely span the space of irreducible characters of the $\mathcal{N} = 4$ $SU(2)$ chiral algebra \cite{Adamovic:2014lra}, where $\mathcal{J}_{[1,1]}$ simply gives the Schur index $\mathcal{I}_{SU(2)}$. There is one logarithmic partner of $\mathcal{J}_{[1,1]}$. Hence $\dim \mathcal{V}_{[1,1]} = 2 = w([1,1])$. On the other hand, $\mathcal{J}_{[2]}$ is a Jacobi form, and hence $\dim \mathcal{V}_{[2]} = 1 = w([2])$.

When $N = 3$ and $\mathbf{n} = (n, -n, 0)$,
\begin{align*}
  & \ \mathcal{I}(\mathbf{n})  \\
  = & \ - \frac{i(b^n - b^{-n})[(b^{\frac{n}{2}} + b^{-\frac{n}{2}}) - (q^{\frac{n}{2}} + q^{-\frac{n}{2}})]}{12 (q^{\frac{n}{2}} - q^{-\frac{n}{2}})^3} \mathcal{J}_{[3]} \\
  & \ - \frac{i}{24} \frac{(b^n + b^{-n})(q^{\frac{n}{2}} + q^{-\frac{n}{2}}) - 2 (b^n + b^{-n})}{(q^{\frac{n}{2}} - q^{-\frac{n}{2}})^3} \mathcal{J}_{[2,1]}\\
  & \ - \frac{i (b^n - b^{-n})}{q^{\frac{n}{2}} - q^{-\frac{n}{2}}}\mathcal{J}_{[1,1,1]} \ .
\end{align*}
The three $\mathcal{J}$ span all $SU(3)$ Wilson-line indices. Under $\Gamma^0(2)$, $\dim \mathcal{V}_{[3]} = 1$, $\dim \mathcal{V}_{[2,1]} = 2$, and $\dim \mathcal{V}_{[1,1,1]} = 3$.

\section{Mixed Hodge polynomials}

The 4D $\mathcal{N} = 4$ $SU(N)$ theory is engineered by compactifying the 6D $(0,2)$ SCFT on a genus-one Riemann surface with a puncture of type $[N - 1, 1]$. When the theory is further compactified on $S^1$, the Coulomb branch $\mathcal{M}_N$ can be described as the following character variety of $PGL(N, \mathbb{C})$, the Langlands dual of $SU(N)$:
\begin{align}
  \mathcal{M}_N
  \coloneqq \Big\{(a, b, x) &\ \in PGL(N, \mathbb{C})^2 \times C \\
  & \ \Big | \ x [a, b] = \operatorname{id}_N \Big\}//PGL(N, \mathbb{C}) \ . \nonumber
\end{align}
Here $C$ denotes a conjugacy class in $PGL(N, \mathbb{C})$ corresponding to the partition $[N-1, 1]$, $[a, b]\coloneqq a b a^{-1} b^{-1}$, and $//PGL(N, \mathbb{C})$ denotes the quotient by conjugation. Supersymmetry endows the Coulomb branch $\mathcal{M}_N$ with a hyperk\"{a}hler structure. A smooth K\"{a}hler structure would imply a standard decomposition of differential forms into $(p, q)$-forms, such that the degree $k$ of a differential form satisfies $k = p + q$. However, the Coulomb branch $\mathcal{M}_N$ is noncompact and singular, giving rise to a nontrivial mixed Hodge structure on its cohomology. Loosely speaking, this structure captures the discrepancy between $k$ and $p + q$. The matching between the mixed Hodge polynomial and the representation theory of the chiral algebra was first observed in \cite{Pan:2024hcz,Pan:2024epf} for the $A_1$ class-$\mathcal{S}$ theories; here, we prove the correspondence concretely for the $\mathcal{N} = 4$ theories.

The compactly supported mixed Hodge polynomial $H_c(\mathcal{M}_N; q, t) = \sum_{i, j}h_c^{i,i;j}(\mathcal{M}_N)q^i t^j$ was computed in \cite{hausel2008mixed}:
\begin{align}
  H_c(\mathcal{M}_N; q, t) = (t\sqrt{q})^{d_N} \mathbb{H}^{[N]}(\sqrt{q}, -\frac{1}{t\sqrt{q}}).
\end{align}
Here $d_N$ is the dimension of $\mathcal{M}_N$:
\begin{equation}
	d_N = 2(N-1) \, ,
\end{equation}
and the polynomial $\mathbb{H}^{[N]}(z, w)$ is generated by
\begin{equation}
  \sum_{N \ge 0} (z-w)^2\mathbb{H}^{[N]}(z, w) T^N
  = \prod_{n \ge 1} \frac{(1 - zw T^n)^2}{(1 - z^2 T^n)(1 - w^2 T^n)} \ .
\end{equation}
From $H_c$, one can construct the Poincar\'{e} polynomial $PH_c$ of the pure part of the cohomology of $\mathcal{M}_N$,
\begin{equation}
  PH_c(\mathcal{M}_N; q) \coloneqq q^{d_N/2} \mathbb{H}^{[N]}(\sqrt{q}, 0) \ .
\end{equation}
The factor $(z-w)^2$ arises because we consider the character variety of $PGL(N,\bbC)$ rather than that of $GL(N,\bbC)$.

Hence, the Poincar\'{e} polynomial $PH_c(\mathcal{M}_N; q)$ can be obtained directly from the contour integral
\begin{equation}
  PH_c(\mathcal{M}_N;q) = \oint \frac{dT}{2\pi i T} T^{-N} \frac{1 - q}{(q;T)} \ .
\end{equation}
Here $(q;T) \coloneqq \prod_{n \ge 0}(1 - q T^n)$ is the $T$-Pochhammer symbol. In other words,
\begin{equation}
  PH_c(\mathcal{M}_N; q) = 
  \sum_{[1^{m_1} \cdots N^{m_N}] \in \mathcal{P}(N)} q^{m_1 + \cdots + m_N - 1}\ .
\end{equation}
Identifying $[1^{m_1} \cdots N^{m_N}]$ with $\boldsymbol{\ell} = [\ell_1, \cdots, \ell_w]$, we have
\begin{equation}
  m_1 + \cdots + m_N = w(\boldsymbol{\ell}) \ ,
\end{equation}
and therefore,
\begin{equation}
  PH_c(\mathcal{M}_N; q) = 
  \sum_{\boldsymbol{\ell} \in \mathcal{P}(N)} q^{w(\boldsymbol{\ell})-1}
  = \sum_{\boldsymbol{\ell} \in \mathcal{P}(N)} q^{\dim \mathcal{V}_{\boldsymbol{\ell}}-1} \ .
\end{equation}
Hence, each term in the polynomial corresponds to an irreducible character of $\mathbb{V}[\mathcal{T}_{SU(N)}]$, while its exponent encodes the number of logarithmic partners through $\dim \mathcal{V}_{\boldsymbol{\ell}}$.

\section{Discussion}

To summarize, this Letter identifies the character space of the $\mathcal{N} = 4$ $SU(N)$ chiral algebra $\mathbb{V}[\mathcal{T}_{SU(N)}]$ by computing the Wilson line index of the 4D theory and systematically identifying the rational-function coefficients that reveal the hidden characters. The irreducible characters are in one-to-one correspondence with the integer partitions $\boldsymbol{\ell}$ of $N$. The logarithmic nature of the algebra implies the existence of logarithmic characters, which are constructed by imposing modular invariance on the character space $\mathcal{V}$. We prove that the structure of $\mathcal{V}$ is encoded in the Coulomb branch by computing the pure part of its mixed Hodge polynomial.

The representation theory of chiral algebras is a rich and challenging subject. Our results show that it is nontrivially encoded in the nonlocal operator spectrum of the 4D theory and can be extracted from exact computations. It would be highly desirable to match the chiral algebra data with the fixed-point varieties of the Coulomb branch on $S^1$, a standard entry in the 4D mirror-symmetry dictionary. Besides line operators, Gukov-Witten-type surface operators and their indices constitute another important subject of study \cite{Gukov:2006jk,Gukov:2008sn}. These operators are also determined by partitions of $N$, suggesting a nontrivial relation to the representation theory of $\mathbb{V}[\mathcal{T}_{SU(N)}]$. It would also be interesting to generalize our results to $\mathcal{N} = 4$ theories with other gauge groups, higher-rank class-$\mathcal{S}$ theories, and Argyres-Douglas theories with exactly marginal deformations \cite{Beem:2023ofp}, thereby establishing concrete relations between the physics of the Higgs and Coulomb branches of 4D $\mathcal{N} = 2$ theories.

\begin{acknowledgments}
The authors would like to thank Tomoyuki Arakawa, Yongchao L\"u, Peng Shan, and Samuel DeHority.
W.Y. is supported by the National Key R\&D Program of China (Grant No. 2025YFA1017400). The work of Y.P. is supported by the National Natural Science Foundation of China (NSFC) under Grant No. 11905301.
\end{acknowledgments}

\bibliography{ref}

\end{document}